\documentclass[twocolumn,twoside,slac_two]{revtex4}
\pdfoutput=1
\usepackage{graphicx}
\usepackage{fancyhdr}
\begin{document}

\def\babar{\mbox{\sl B\hspace{-0.4em} {\footnotesize\sl A}\hspace{-0.37em}  \sl B\hspace{-0.4em} {\footnotesize\sl A\hspace{-0.02em}R}}}

\title{Lepton Universality Test in ${\bf \Upsilon}$(1S) decays at \emph{{\Large B}\hspace{-0.4em} {\large A}\hspace{-0.37em}   {\Large B}\hspace{-0.4em} {\large A\hspace{-0.02em}R}}}

\author{Elisa GUIDO (on behalf of the  \babar\ Collaboration)}
\affiliation{Universit\`a degli Studi di Genova e INFN - Sezione di Genova, Italy}

\begin{abstract}
Using a sample of 122 million $\Upsilon(3S)$ decays collected with the {\small\sl B\hspace{-0.4em}} {\footnotesize\sl A}\hspace{-0.37em} {\small\sl B\hspace{-0.4em}} {\footnotesize\sl A\hspace{-0.02em}R} detector at the PEP-II asymmetric energy collider at the SLAC National Accelerator Laboratory, we measure the ratio $R_{\tau\mu}=BR(\Upsilon(1S)\to\tau^+\tau^-)/BR(\Upsilon(1S)\to\mu^+\mu^-)$; the measurement is intended as a test of lepton universality and as a possible search for a light pseudoscalar Higgs boson in Next to Minimal Supersymmetric Standard Model (NMSSM) scenarios. Such a boson could appear in a deviation of the ratio $R_{\tau\mu}$ from the Standard Model expectation, that is 1, except for small lepton mass corrections. The analysis exploits the decays $\Upsilon(3S) \to \Upsilon(1S)\pi^+\pi^-$, $\Upsilon(1S) \to l^+l^-$, where $l=\mu, \tau$. 
\end{abstract}

\maketitle

\thispagestyle{fancy}


\section{Introduction}

In the Standard Model (SM), the couplings between gauge bosons and leptons are independent of lepton flavor. Therefore, the branching ratio $BR(\Upsilon(1S)\to l^+l^-)$ should basically not depend on the lepton $l$ considered. The quantity:

\begin{equation}
R_{ll'}=\frac{BR(\Upsilon(1S)\to l^+l^-)}{BR(\Upsilon(1S)\to l'^+l'^-)},
\end{equation}
where $l, l' = e, \mu, \tau$, is thus expected to be very near to one, the discrepancy from unity being due to small lepton-mass effects (for instance, the greatest correction should turn out in $R_{\tau\mu}\sim0.992$).

Beyond the SM, deviations of $R_{ll'}$ from this expectation are possible. In particular, in the Next to Minimal Supersymmetric Standard Model (NMSSM)~\cite{ref:Higgs} scenario, there is the hypothesis of existence of a light pseudo-scalar Higgs boson, namely $A^0$, which could be escaped to LEP bounds~\cite{ref:LEP1, ref:LEP2}.
For instance, considering the leptonic decays of the $\Upsilon(1S)$, such a boson could mediate the decay chain of the resonance in one of the following modes~\cite{ref:miguelangel1, ref:miguelangel2}:

\begin{equation}
\Upsilon(1S)\to A^0\gamma, A^0\to l^+l^-\label{eqn:a0_nomix}
\end{equation}
or
\begin{equation}
\Upsilon(1S)\to\eta_b(1S)\gamma, \eta_b(1S)\to A^0\to l^+l^-.\label{eqn:a0_etabmix}
\end{equation}
The latter decay implies a mixing between $\eta_b(1S)$, i.e. the ground state of the bottomonium family, recently discovered by \babar\ ~\cite{ref:eta_b3S}, and which does not have the right quantum numbers to decay in a pair of leptons, and the light Higgs $A^0$.
If the photon was present but remained undetected because of its softness, the lepton pair would be ascribed to the $\Upsilon(1S)$. This would result in a deviation of $R_{ll'}$ from the SM expectation, that is in a lepton universality violation. Similar considerations hold on for the higher-mass $\Upsilon$ resonances too.
The coupling between $A^0$ and the lepton being proportional to the lepton mass, this effect should be more evident when one of the leptons is a $\tau$. Theoretical calculations~\cite{ref:miguelangel1, ref:miguelangel2} foresee effects up to an order of $\sim10\%$.

The CLEO Collaboration has studied the decays $\Upsilon(nS)\to\mu^+\mu^-$, $\tau^+\tau^-$, with $n=1,2,3$, using a sample of about 1 fb$^{-1}$ of data taken at each $\Upsilon(nS)$ peak, corresponding to about $10^7$ $\Upsilon$ resonances. In particular, for the $\Upsilon(1S)$, they measured the branching fraction ratio:

\begin{center}
$R_{\tau\mu}=1.02\pm0.02$(stat.)$\pm0.05$(syst.),
\end{center} 
deriving:

\begin{center}
$B(\Upsilon(1S)\to\eta_b\gamma)\times B(\eta_b\to A^0\to\tau^+\tau^-)<0.27\%$
\end{center}
at the $95\%$ of confidence level~\cite{ref:cleoresults}.

\section{The \emph{{\large B}\hspace{-0.4em}  {\small A}\hspace{-0.37em}   {\large B}\hspace{-0.4em} {\small A\hspace{-0.02em}R}} detector and the data sample}

We present a measurement of $R_{\tau\mu}$ in $\Upsilon(1S)$ decays exploiting the  sample of $\Upsilon(3S)$ decays collected by the \babar\ detector at the PEP-II asymmetric $e^+e^-$ collider
 at the SLAC National Accelerator Laboratory. The available statistics ($\sim 28$ fb$^{-1}$, corresponding to $(121.8\pm1.2)\times10^6$ events) represents the largest sample of $\Upsilon(3S)$ decays ever collected.
The 2.4 fb$^{-1}$ of data collected 30 MeV below the $\Upsilon(3S)$ resonance ({\it off-resonance} sample) are also used for background studies.

The \babar\ detector is described in detail elsewhere~\cite{ref:babar}. Charged-particle momenta are measured in a tracking system consisting of a five-layer double-sided silicon vertex tracker (SVT) and a 40-layer central drift chamber (DCH), both situated in a 1.5-T axial magnetic field. Charged-particle identification is based on the $dE/dx$ measured in the SVT and DCH, and on a measurement of the photons produced in the synthetic fused-silica bars of the ring-imaging Cherenkov detector. A CsI(Tl) electromagnetic calorimeter is used to detect and identify photons and electrons, while muons are identified in the instrumented flux return of the magnet. We use the GEANT~\cite{ref:geant} software to simulate interactions of particles traversing the \babar\ detector, taking into account the varying detector conditions and beam backgrounds.

\section{Analysis strategy}

The signal events ($\Upsilon(1S)\to\mu^+\mu^-$ or $\tau^+\tau^-$) are tagged through the transition $\Upsilon(3S)\to\Upsilon(1S)\pi^+\pi^-$, whose branching fraction is $BR(\Upsilon(3S)\to\Upsilon(1S)\pi^+\pi^-)\sim4.8\%$~\cite{ref:PDG}. We consider all the decays of $\tau$ to one charged track plus undetected neutrals: this choice leads to a very clear environment for our study, since both the typologies of  final states ($\mu^+\mu^-$ and $\tau^+\tau^-$) contain four charged tracks, allowing for a better handling of backgrounds and a good trigger performance. 

Because of the presence of neutrinos in the final state, the $\Upsilon(3S)\to\Upsilon(1S)\pi^+\pi^-$, $\Upsilon(1S)\to\tau^+\tau^-$ decays are not fully reconstructed, and suffer from a larger contamination from background with respect to the $\Upsilon(1S)\to\mu^+\mu^-$ events. For this reason the event selections are mostly different for the two categories of final states.

The event selections are tuned using Monte Carlo simulated samples.
The main sources of background are $e^+e^-\to q\bar q$ (with $q=u, d, s, c$) and $e^+e^-\to\tau^+\tau^-$ continuum events, Bhabha events and generic $\Upsilon(1S)$ decays; the latter produce a peaking yield.
The two typologies of signal events ($\mu^+\mu^-$ and $\tau^+\tau^-$ final states) are separated thanks to a cut on the difference between the visible and the reconstructed energy, calculated in the $e^+e^-$ center of mass frame (CM): indeed, no missing energy is expected in $\Upsilon(1S)\to\mu^+\mu^-$ events, that are completely reconstructed, while an amount of missing energy is actually present in $\Upsilon(1S)\to\tau^+\tau^-$ decays.

\begin{figure}[h]
\centering
\includegraphics[width=80mm]{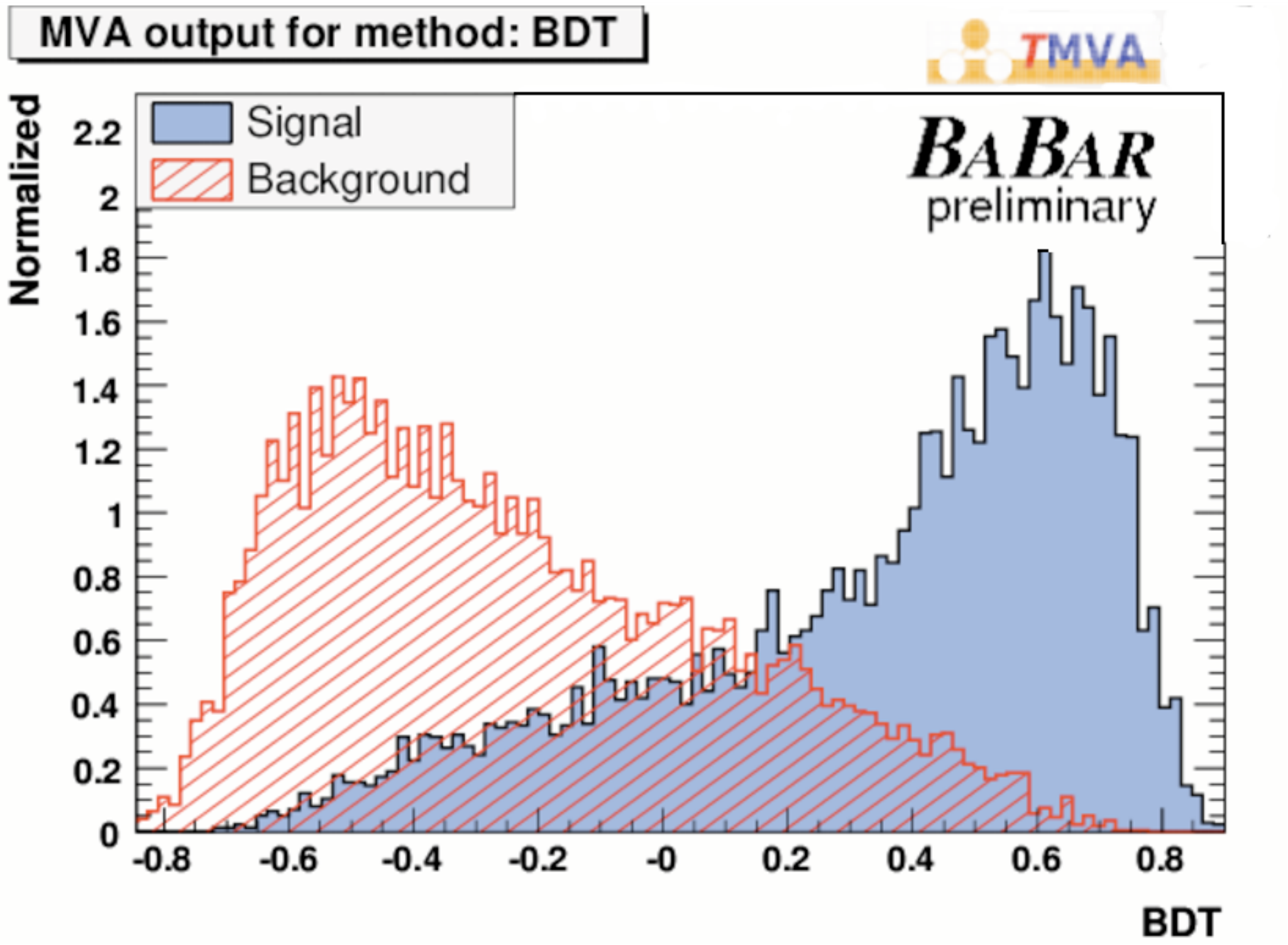}
\caption{ Output variable resulting from the multivariate analysis method, exploiting a boosted decision tree as a 
discriminator tool, for signal Monte Carlo simulation (solid blue) and for {\it off-resonance} data (striped red).} \label{fig:mva}
\end{figure}

Moreover, to further reduce the background in the $\tau^+\tau^-$ final states, a boosted decision tree~\cite{ref:BDT} exploiting several shape and kinematic variables is used. The performance of the classifier is assessed using signal Monte Carlo simulations and {\it off-resonance} data. After the application of the multivariate approach, the separation between signal and background is sensitively improved, as visible in Figure~\ref{fig:mva}. 

Additional cuts are applied for the $\mu^+\mu^-$ selection on the cosine of the angle between the two-lepton candidates calculated with respect to the CM frame, for the $\tau^+\tau^-$ selection on the difference in the energy of the $\Upsilon(3S)$ and the $\Upsilon(1S)$, and for both the selections on the momentum of each of the two pions, as well as of the pair of them.

The signal extraction efficiency is $\epsilon_{\mu\mu}\sim45\%$ ($\epsilon_{\tau\tau}\sim17\%$) for $\mu^+\mu^-$ ($\tau^+\tau^-$) final states.

\section{Signal extraction}

The signal extraction is performed using an extended and unbinned maximum-likelihood fit; the likelihood for the $\Upsilon(1S)\to\mu^+\mu^-$ events is 2-dimensional and based on the variables $\Delta M=M_{\Upsilon(3S)}-M_{\Upsilon(1S)}$, that is the difference between the invariant masses of the two $\Upsilon$ resonances involved in the decay chain, and $M_{\mu^+\mu^-}$, the invariant dimuon mass.
Instead, for the $\Upsilon(1S)\to\tau^+\tau^-$ events, the signal extraction cannot be performed in the same way: the partial reconstruction of the final state, due to the presence of neutrinos from the $\tau$'s decays, means that neither $\Delta M$ nor the dilepton invariant mass can be used to discriminate between signal and background. Therefore, we extract the signal yield performing a 1-dimensional fit to the recoil mass of the pair of pions involved in the decay chain, defined as:

\begin{equation}
M_{\pi^+\pi^-}^{reco} = \sqrt{s+M^2_{\pi\pi}-2\sqrt{s}\sqrt{M_{\pi\pi}^2+p_{\pi\pi CM}^2}}, 
\end{equation}
where $\sqrt{s}$ indicates the nominal CM energy~\cite{ref:PDG}, $M_{\pi\pi}$ indicates the invariant mass of the two pions, and $p_{\pi\pi CM}$ is the momentum of the two pions calculated in the CM frame.
Because this variable is related only to the $\Upsilon(3S)\to\pi^+\pi^-\Upsilon(1S)$ transition, it cannot distinguish between $\Upsilon(1S)\to\tau^+\tau^-$ events and other $\Upsilon(1S)$ decays or the Higgs-mediated events of Eq.~\ref{eqn:a0_nomix} and Eq.~\ref{eqn:a0_etabmix}. 
We use a simulated sample of inclusive $\Upsilon(1S)$ decays to verify that this background becomes negligible when the event selection is applied. Additional tests are performed with the data sample to prove this conclusion, as discussed in detail below. An enhancement of the $\Upsilon(1S)\to\tau^+\tau^-$ yield would thus constitute a possible signal of New Physics.

The two fits are performed simultaneously, and the value of  $R_{\tau\mu}$ is returned.  The likelihood is written as

\begin{equation}
{\cal L}_{ext}  =  \frac{e^{-N'}(N')^N}{N!} \prod_{i=1}^{N} {\cal P}_i
\label{eq:likelihood}
\end{equation}
where ${\cal P}_i$ is:
\begin{eqnarray}
{\cal P}_i & \equiv & N_{\mu} {\cal P}_i^{\mu\mu}(\Delta M, M_{\mu^+\mu^-}) + \nonumber \\ 
 & + & N_{bkg\mu} {\cal P}_i^{bkg\mu}(\Delta M, M_{\mu^+\mu^-}) +\nonumber \\
 & + & \frac{\epsilon_{\tau\tau}}{\epsilon_{\mu\mu}}N_{\mu}R_{\tau\mu}{\cal P}_i^{\tau\tau}(M_{\pi^+\pi^-}^{reco}) +\nonumber \\
 & + & N_{bkg\tau}{\cal P}_i^{bkg\tau}(M_{\pi^+\pi^-}^{reco}) \label{eqn:LH}
\end{eqnarray}
and $N_\mu$, $N_{bkg\mu}$ and $N_{bkg\tau}$ indicate the number of the signal events extracted in the $\mu^+\mu^-$ sample, the number of background events extracted in the $\mu^+\mu^-$ sample and the number of background events extracted in the $\tau^+\tau^-$ sample, respectively.

In the fit, the probability density functions (PDFs) for the signal components in $\Upsilon(1S)\to\mu^+\mu^-$ sample are taken from a subset of  data (about one tenth of the complete available statistics, which is discarded from the eventual result in order to avoid any possible bias); the $\Delta M$ distribution is described by a triple Gaussian function with uncommon means, while the $M_{\mu^+\mu^-}$ distribution is modeled by an analytical function approximating a Gaussian with mean value equal to $\mu$ and different left and right widths $\sigma(L,R)$, plus asymmetric non gaussian tails $\alpha(L,R)$, defined as:

\begin{equation}
{\cal F}(x) = exp\Big\{ -\frac{(x-\mu)^2}{2\sigma^2(L,R)+\alpha(L,R)(x-\mu)^2} \Big\}. \label{eqn:Cruijff}
\end{equation}

In $\Upsilon(1S)\to\tau^+\tau^-$ events, the PDF for the signal component is taken from the data sample $\Upsilon(1S)\to\mu^+\mu^-$ sample (where the dipion recoil mass is not used as a variable for the signal extraction and is therefore independent of the fit procedure); the $M_{\pi^+\pi^-}^{reco}$ distribution is described by the function defined in Eq.~\ref{eqn:Cruijff}. All the parameters of the signal PDFs are fixed in the fit, while the background shapes (flat functions in $\mu^+\mu^-$ sample and polynomial function of first order in $\tau^+\tau^-$ sample) are floated in the fit. We use the {\it off-resonance} sample to verify that the analytic description of the background shapes is appropriate.

\begin{figure*}[t]
\centering
\includegraphics[width=80mm]{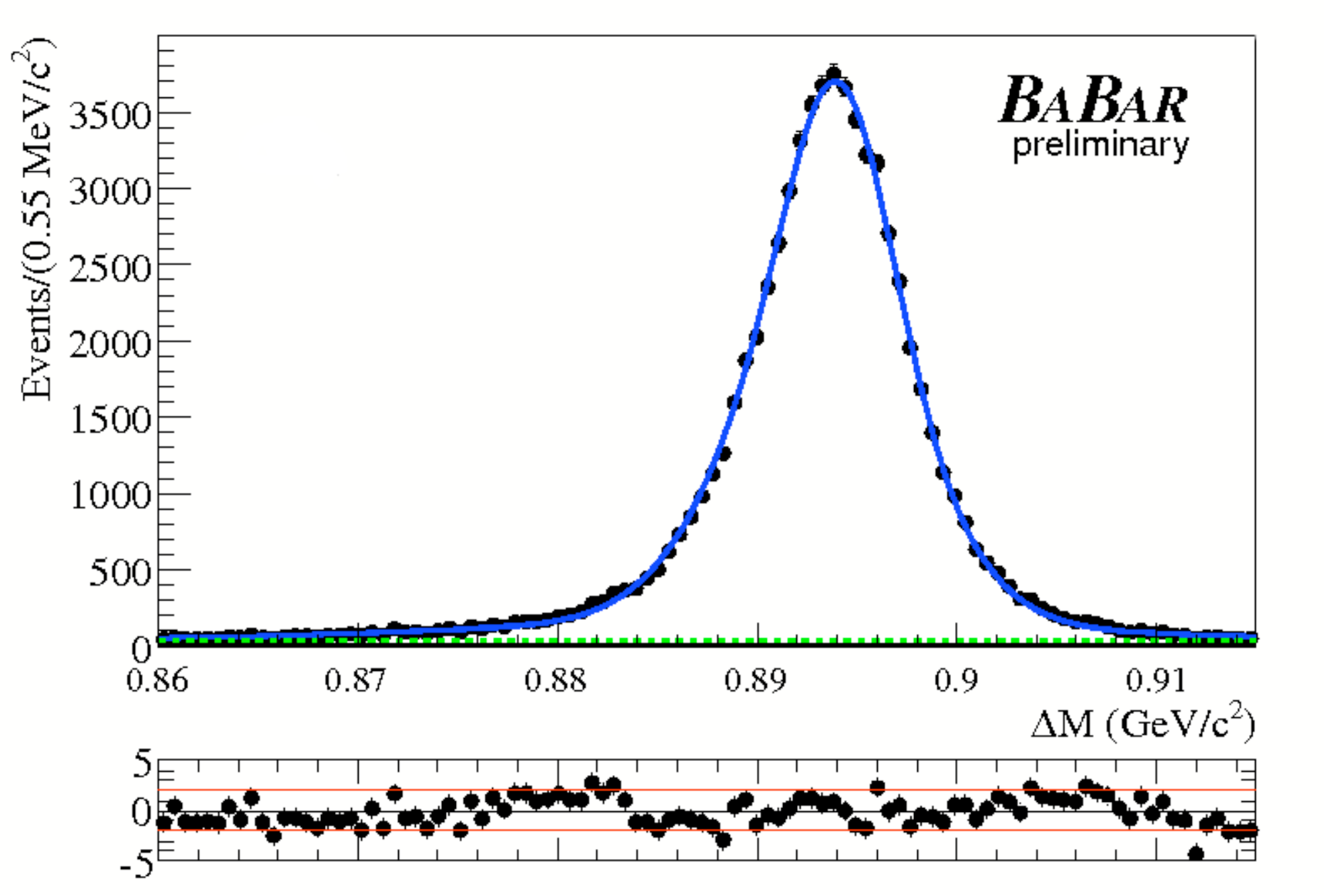}\includegraphics[width=80mm]{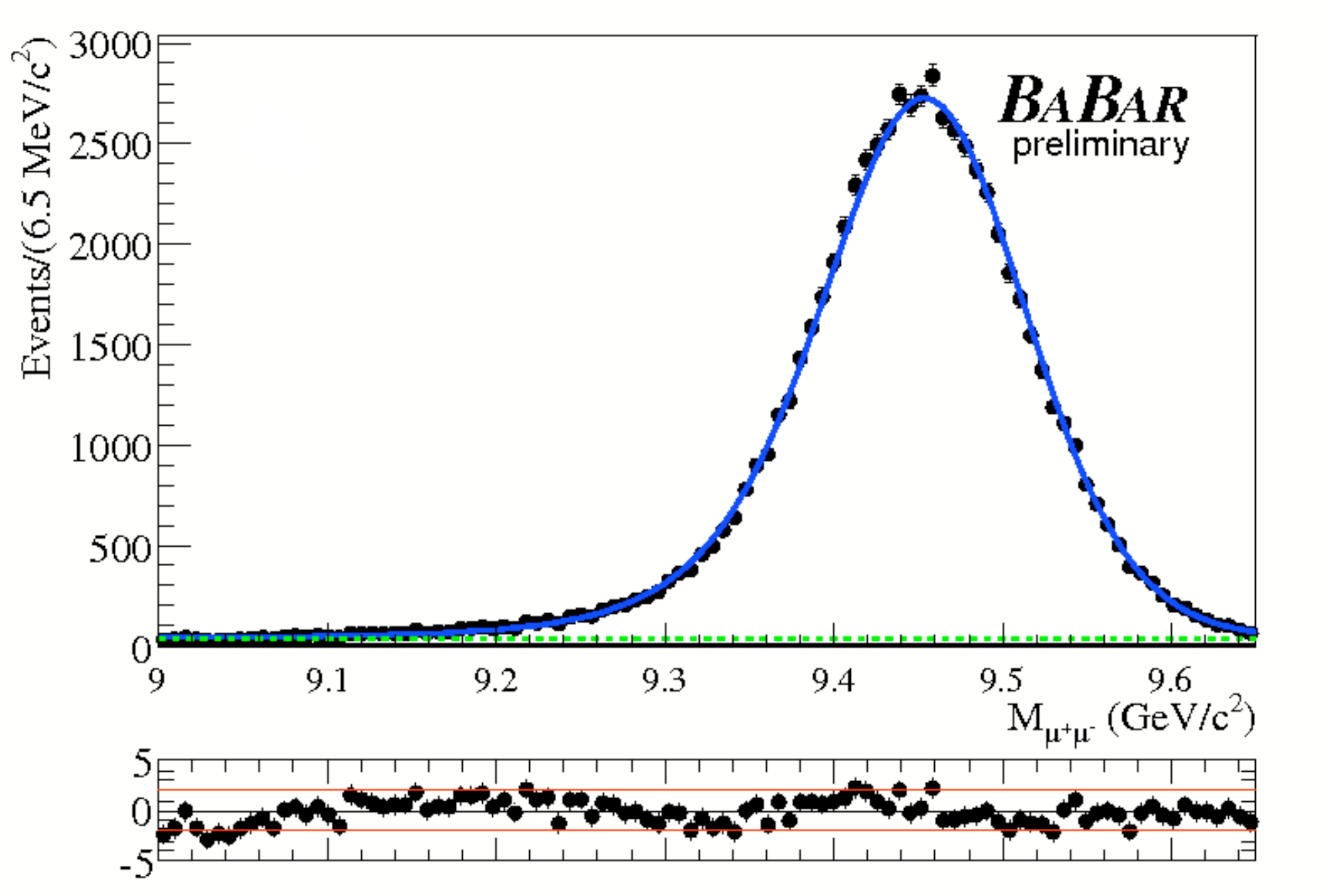}\\
\includegraphics[width=80mm]{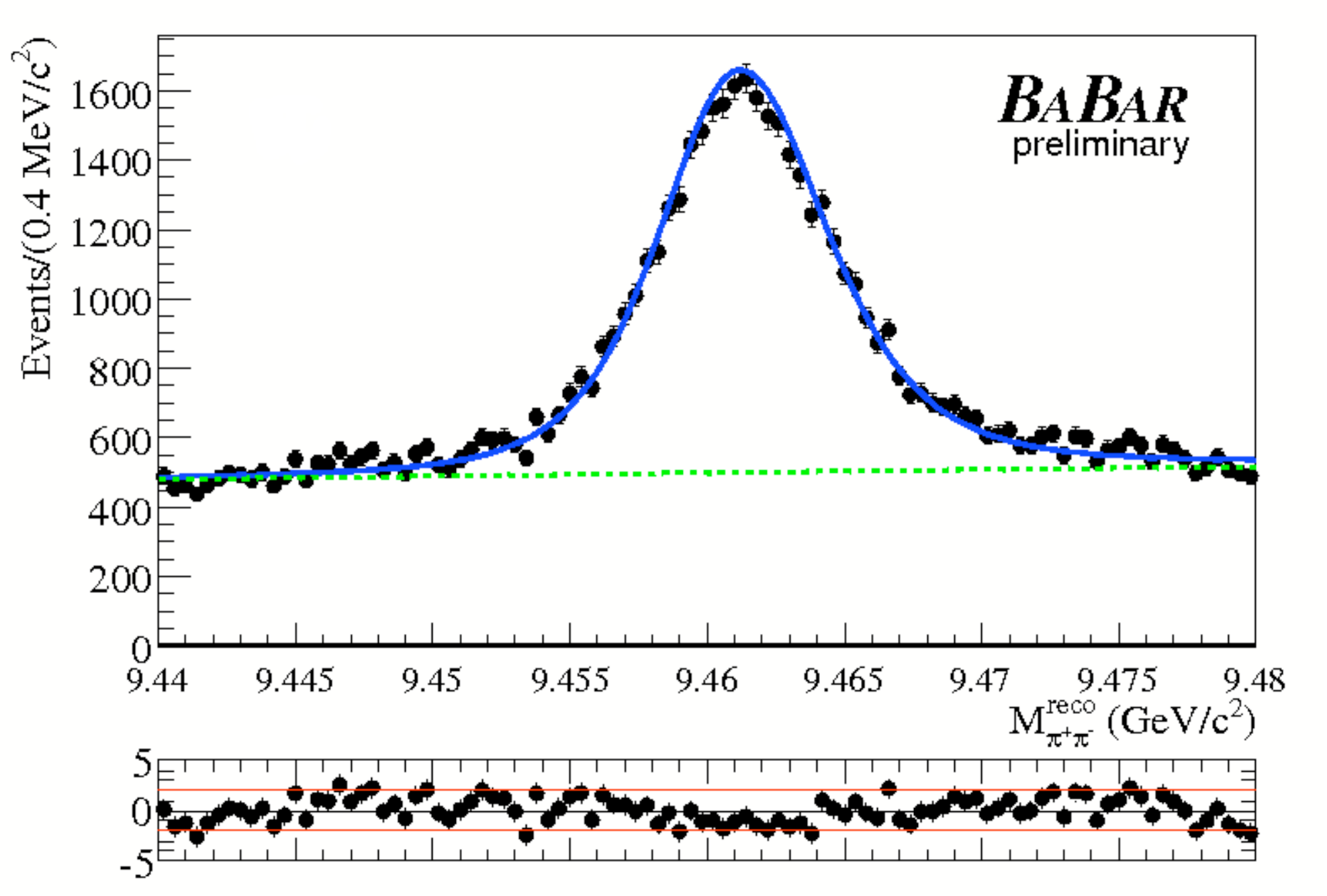}
\caption{Results of the fit procedure for the $\Delta M$ distribution (top left), for the $M_{\mu^+\mu^-}$ distribution (top right) and for the $M_{\pi^+\pi^-}^{reco}$ distribution (bottom). In each plot the dashed green line represents the background shape, while the solid blue line is the sum of signal and background contributions. The residuals of the fit are shown under each plot, with the solid red lines indicating the 2$\sigma$ level of agreement.} \label{fig:fit}
\end{figure*}

The result of the simultaneous fit procedure is shown in Figure~\ref{fig:fit}, while Table~\ref{tab:fit} shows the results for the signal and background yields, along with the results of the $\chi^2/d.o.f.$ tests, made to compare the data distribution with the curve chosen to describe it (where $d.o.f.$ indicates the number of degrees of freedom, namely equal to the number of histogram bins minus the number of parameters fixed in the PDF).

\begin{table}[t]
\begin{center}
\caption{Results of the simultaneous fit: 
signal and background yields; $R_{\tau\mu}$; output of $\chi^2/d.o.f.$ tests. The results are only statistical; no systematic correction is applied yet.}
\begin{tabular}{|c|c|}  
\hline
N$_\mu$     & $60192 \pm 269$ \\
N$_{bkg\mu}$     & $3136 \pm 92$ \\
N$_{bkg\tau}$     & $44023 \pm 253$ \\
$R_{\tau\mu}$     & $1.006 \pm 0.010$ \\
\hline
$\chi^2(\Delta M)/d.o.f.$ & 176/92\\
$\chi^2(M_{\mu^+\mu^-})/d.o.f.$ & 133/95\\
$\chi^2(M_{\pi^+\pi^-}^{reco})/d.o.f.$ & 158/95\\
\hline
\end{tabular}
\label{tab:fit}
\end{center}
\end{table}

\section{Systematic uncertainties}

In the measurement of $R_{\tau\mu}$, several systematic uncertainties
are common to the numerator and
denominator of the ratio, and therefore cancel out. These include uncertainties on the luminosity, the
$\Upsilon(3S)$ production cross section, the $\Upsilon(3S)\to
\Upsilon(1S)\pi^+\pi^-$ branching fractions, as well as uncertainties from
differences between data and simulation in track reconstruction
efficiencies and the common event selection.

The residual systematic uncertainties that do not cancel arise from:
\begin{itemize}
\item systematic uncertainty related to the event selection efficiency;
\item differences between data and simulation in the particle
  identification (PID) efficiency;
\item differences between data and simulation in trigger and
  background filter efficiency;
\item imperfect knowledge of the signal and background shapes used in
  the fit to extract the number of signal events on data.
\end{itemize}

The systematic uncertainty due to the event selection is evaluated as the discrepancy between data and simulation in the variation of the efficiency produced by removing one cut at a time from the complete event selection. 
This procedure is performed for those selection criteria which are not common between the two channels (and hence do not cancel) and also for the selection criteria which use different values of the same variables (which partially cancel).
The systematic error is $1.5\%$, mainly due to the effect of the cut applied to the output variable of the multivariate technique used for $\Upsilon(1S)\to\tau^+\tau^-$ event selection.

The systematic uncertainty related to the difference between data and simulation in the PID efficiency only applies to $\Upsilon(1S)\to\mu^+\mu^-$ events, and is estimated by exploiting two independent samples: one where both leptons are required to be identified as $\mu$ leptons, and another where exactly one final charged track is a muon. We determine the relative inefficiency for requiring the second sample instead of the first one, both on data and on simulation; the ratio of the two results gives an efficiency correction of  1.023  and  a related systematic error of $0.6\%$.

The systematic uncertainties due to the differences between data and simulation in trigger and background digital filters (BGF) efficiency are small both in $\Upsilon(1S)\to\mu^+\mu^-$ and in $\Upsilon(1S)\to\tau^+\tau^-$ events, and they cancel at least in part in the ratio. These values are separately estimated for each channel by comparing the efficiency for the signal extraction with and without the trigger requirements (and similarly with or without the BGF), both on data and on simulation. Concerning the trigger, a correction of 1.020 is needed for the $\epsilon_{\tau\tau}$ efficiency, together with a systematic uncertainty of $0.10\%$ for $\Upsilon(1S)\to\tau^+\tau^-$ events, while a systematic uncertainty of $0.18\%$ is quoted for $\Upsilon(1S)\to\mu^+\mu^-$ events. The contribution given by BGF efficiency to the total uncertainty is negligible.

The imperfect knowledge of the signal and background shapes used in the signal extraction procedure is taken into account considering:

\begin{itemize}
\item the systematic uncertainty resulting from fixing the parameters of the functional forms describing the signal distributions (up to $1.7\%$); this error is estimated by repeating the fit procedure changing of $\pm 1\sigma$ all the parameters fixed, one at a time, and summing in quadrature the discrepancies between the nominal result (Table~\ref{tab:fit}) and the one obtained in each of the modified configurations;
\item the systematic uncertainty due to the choice of the background parameterization for each distribution used in the fit procedure: it is evaluated by using alternative parameterizations describing the background shape (polynomial functions of first order for $\Delta M$ and $M_{\mu^+\mu^-}$, flat function and polynomial function of second order for $M_{\pi^+\pi^-}^{reco}$) and summing in quadrature the discrepancies between the nominal result and the one obtained in each of the modified configurations (up to $0.28\%$);
\item the systematic discrepancy between the sample used to determine the parameters of each distribution. For $\Upsilon(1S)\to\mu^+\mu^-$ events, a sub-sample of data themselves is used, leading to no expected bias and therefore without a systematic assigned. For $\Upsilon(1S)\to\tau^+\tau^-$ events, the $\mu^+\mu^-$ sample is exploited, and an estimate of the systematic discrepancy between the two samples is needed. It is evaluated by re-weighting the parameters for the $M_{\pi^+\pi^-}^{reco}$ distribution with the ratio between the parameters obtained comparing the fit result on the Monte Carlo samples for both the leptonic $\Upsilon(1S)$ decay channels. We estimate this contribution to be about $0.10\%$. Moreover, in order to consider further discrepancies possibly not taken into account by simulations, an additional systematic uncertainty is estimated by repeating the fit procedure while allowing the global width of the function defined in Eq.~\ref{eqn:Cruijff} (which describes the $M_{\pi^+\pi^-}^{reco}$ shape for the signal) to float; the result is a discrepancy of $0.05\%$ in $R_{\tau\mu}$. The total contribution is therefore $0.11\%$.
\end{itemize}

Finally the statistics of the Monte Carlo simulated samples used gives a contribution to the systematic uncertainty of less than $0.1\%$ in both the leptonic final states.

Additional studies have been performed in order to check the possible background contribution to the $\tau^+\tau^-$ sample after applying the selection. The signal extraction is repeated requiring the two $\tau$ leptons in the $\Upsilon(3S)\to\Upsilon(1S)\pi^+\pi^-\to\tau^+\tau^-\pi^+\pi^-$ decay chain to be detected in the final state $e^+\mu^-$ or $\mu^+ e^-$. 
The central values for $R_{\tau\mu}$ obtained with the two procedures on the control sub-sample of the complete dataset (already used in other steps of the analysis and discarded from the final result) are compatible: $R_{\tau\mu}^{nom}=1.006\pm0.029$(stat.) and $R_{\tau\mu}^{PID}=0.999\pm0.044$(stat.), where the first value indicates the result obtained with the nominal procedure and the second one the value obtained requiring $\tau^+\tau^-\to e^\pm\mu^\mp +$ neutrinos. This further confirms that the peaking background due to generic $\Upsilon(1S)$ decays is indeed negligible.

The summary of the systematic uncertainties is shown in Table~\ref{tab:syst}. The total systematic error, obtained by summing in quadrature all the contributions, is estimated to be $2.4\%$.

\begin{table}[t]
\begin{center}
\caption{Summary of systematic errors and correction to the efficiency for $\mu^+\mu^-$ channel (first column) and for $\tau^+\tau^-$ channel (second column). The total systematic uncertainty is also shown.}
\begin{tabular}{|c|c|c|}
\hline
{\bf Systematic error:} & $\mu^+\mu^-$ & $\tau^+\tau^-$\\
\hline 
Event selection        & \multicolumn{2}{c|}{$1.5\%$}\\
PID    &  $0.6\%$  & --- \\
Trigger &  $0.18\%$  & $0.10\%$  \\
BGF & negl. & negl. \\
PDF parameters  &\multicolumn{2}{c|}{$1.7\%$}\\
Background PDFs  &\multicolumn{2}{c|}{$0.28\%$}\\
Agreement $\mu^+\mu^-$ vs. $\tau^+\tau^-$ in $M_{\pi^+\pi^-}^{reco}$ & --- & $0.11\%$\\
MC statistics & $0.08\%$ & $0.09\%$\\
\hline
\hline
TOTAL  &\multicolumn{2}{c|}{$2.4\%$}\\
\hline
\end{tabular}
\label{tab:syst}
\end{center}
\end{table}

\section{Results}

Including all the systematic corrections, the preliminary result obtained by \babar\ is:
\[
R_{\tau\mu}=1.009 \pm 0.010(stat.) \pm 0.024(syst.).
\]
Therefore we do not observe any significant deviation of the ratio $R_{\tau\mu}$ from the SM expectation.
With the present result the \babar\ Collaboration achieves a greater level of precision with respect to the previous best value~\cite{ref:cleoresults}, improving both the statistical and the systematic uncertainties.

This result is still preliminary, since a further improvement in the systematic precision is on going. The finalization of the analysis is expected for soon.

\bigskip 


\end{document}